\documentstyle[12pt]{article}
\begin{document}
\begin{titlepage}
\title{
{\LARGE {\bf General relativity reformed to
a genuine Yang-Mills
gauge theory for gravity}}}
\author{
{\normalsize {\bf Nikolaos A. Batakis\thanks
{On leave from the  Dpt of Physics, Univ. of Ioannina,
GR-45110 Ioannina, Greece.
Email: nbatakis@cc.uoi.gr  }}}\\ 
{\normalsize {CERN-Theory Division, CH-1211 Geneva 23, Switzerland}}}
\date{}
\maketitle
\begin{abstract}
\begin{sloppypar}
\normalsize

\noindent
The theory of general relativity is reformed to a genuine Yang-Mills 
gauge theory of the Poincar\'e group for gravity. Several pathologies
of the conventional theory are thus removed, but not every GR 
vacuum satisfies the Y-M equations. The sector of GR solutions which
survive is fully classified and it is found to include the 
Schwarzschild black hole. Two other solutions presented here have 
no GR counterpart and they describe expanding Friedmann universes 
with torsion which vanishes only asymptotically. They are discussed 
along with novel theoretical possibilities, such as a well-defined 
energy-momentum tensor for the gravitational field, and novel 
perspectives for unification and quantization.

\vspace{1cm}
\noindent

\end{sloppypar}
\end{abstract}
\addtolength{\baselineskip}{.3\baselineskip}
\end{titlepage}
\newpage 

\section{Introduction}

It is quite amazing that shortly after the final
formulation of Einstein's theory of general relativity, 
Elie Cartan had essentially supplied the geometric 
framework for a genuine Yang-Mills gauge 
theory of gravity (GGG)\cite{1}.
Following the early and subsequent attempts
(cf. \cite{2},\cite {3} for reviews) the existence of a 
GGG, notably one based on the Poincar\'e 
group ${\cal P}=ISO(1,3)$, has rather been taken for
granted \cite {3}\cite{4}. As far as the present author 
is aware, no actual example or model for a GGG has been
given, a fact related not only to objective difficulties,
but apparently also to certain misconceptions which 
erroneously attribute gauge-theoretic aspects to GR 
and vice-versa. Here, we will particularly need to 
clarify and diggress in the following:
(i) The nature of gauge, isometry and holonomy
transformations, after the gauge fields have been 
soldered onto the space they inhabit.
(ii) The relation of these transformations to what is 
actually lost by the (usual) supression of the 
translational degrees of freedom when gauging ${\cal P}$.
(iii)  The identification of the quadratic-in-cuvature 
part of the action which remains a topological invariant 
when the torsion ${\cal T}$ does not vanish, and the 
emergence of linear-in-${\cal R}$ and quadratic in 
${\cal T}$ terms.
(iv) The reform of GR as an acceptable GGG.

The above also outline our motivation and our approach  
for the genuine (meaning: straigtforward, 
real, unsuppressed) Y-M gauge theory of ${\cal P}$,
presented in sections 2,3 and shown to be free of certain 
pathologies of GR. In section 4, the sector of GR vacua
which survive (namely satisfy the Y-M equations) is fully
classified and shown to include the Schwarzschild space-time, 
and thus the observational backing of the latter. 
Two other solutions found describe expanding Friedmann 
universes and have no GR counterparts. They are discussed in 
section 5, along with the rest of our findings. Our notation 
and terminology generally follows that of \cite{4}.

\section{Gauging the Poincar\'e group}

The generators $\{P_a,M_{aa'}\}$ of ${\cal P}$
satisfy the well known commutation relations
\begin{equation}
[P_a,P_b]=0,\;\;
[M_{aa'},P_b]=2\delta^c_{[a}g_{a']b}P_c,\;\;
[M_{aa'},M_{bb'}]=4\delta^c_{[a} g_{a'][b}\delta^{c'}_{b']}M_{cc'}.
\label{pcr} 
\end{equation}
They are assigned, respectively, to
translations and Lorentz $SO(1,3)$ rotations in
Minkowski space-time $M^4_{\circ}$ serving as a
representation space. ${\cal P}$ acts on $M^4_{\circ}$ as 
a maximal group of isometries which preserve
the standard flat (Minkowski) metric
\begin{equation}
g_{ab}=\mbox{diag}(-1,1,1,1).
\label{m}
\end{equation}
The 1-form valued potential
\begin{equation}
{\cal A}=\theta^aP_a+\frac{1}{2}\omega^{ab}M_{ab},
\label{A}
\end{equation}
inhabits as a gauge field some target
space-time $M^4$, whereupon soldered defines the
vierbein $\theta^a$ and connection $\omega^{ab}$
structure of $M^4$. This construction will be
discussed in more detail shortly.
For the moment we note that
the field strength ${\cal F}$ and 
the covariant derivative ${\cal D}$ are defined 
in terms of ${\cal A}$ as usual
\begin{equation}
{\cal T}^aP_a+\frac{1}{2}{\cal R}^{ab}M_{ab}=
{\cal F}:= d{\cal A}+\frac{1}{2}[{\cal A},{\cal A}],\;\;
\;\;{\cal D}{\cal F}:= d{\cal F}+[{\cal A},{\cal F}]=0.
\label{F}
\end{equation} 
The suggestive identifications for the components
of ${\cal F}$ reflect the emergence of
precisely Cartan's structure equations for the
torsion and curvature 2-forms as  
\begin{equation}
{\cal T}^a:=D\theta^a:=d\theta^a+\omega^a_{\cdot b}\wedge 
\theta^b,\;\;\;\;\;
{\cal R}^a_{\cdot b}:=d\omega^a_{\cdot b}+
\omega^a_{\cdot c}\wedge \omega^c_{\cdot b}, 
\label{c0}
\end{equation}
which automatically satisfy the Bianchi identities
\begin{equation}
D{\cal T}^a={\cal R}^a_{\cdot b}\wedge \theta^b,
\;\;\;\;\;
D{\cal R}^a_b=0. 
\label{bi}
\end{equation}

The Y-M action is
\begin{equation}
{\cal I}_{YM}=\int_{M^4}\mbox{tr}\left(
{\cal F}\wedge\ast{\cal F}\right)=\frac{1}{2}\int_{M^4}
{\cal R}^{ab}\wedge\ast{\cal R}_{ab}
=\frac{1}{4}\int_{M^4}R^{abcd}R_{abcd}\;\eta,
\label{I}
\end{equation}
where the $\ast$ duality is with respect to the metric 
(\ref{m}), the functions $R^a_{bcd}$ specify the components 
of ${\cal R}^a_b$ in the $\theta^a$ frame, and $\eta$ is the 
invariant volume 4-form. The Bianchi identity and (from 
variation of ${\cal A}$) the vacuum Y-M equations  are 
\begin{equation}
{\cal D}{\cal F}=0,\;\;\;\;\;\;{\cal D}\ast{\cal F}=0.
\label{ym}
\end{equation} 
Of these, the first is precisely eqivalent to the Bianchi 
identities (\ref{bi}), while the second may be
written in tems of the components of ${\cal F}$ as
\begin{equation}
D\ast{\cal T}^a=\ast{\cal R}^a_{\cdot b}\wedge \theta^b,
\;\;\;\;\;
D\ast{\cal R}^a_b=0. 
\label{ymc}
\end{equation} 
These field equations may be integrated 
to provide vacuum configurations
according to prescribed boundary or asymptotic conditions.
One may also look for (anti-)self-dual solutions, namely with 
\begin{equation}
{\cal F}=\pm\ast{\cal F}\;\;\;\;\;\;\leftrightarrow\;\;\;\;\;\;
{\cal T}^a=\pm\ast{\cal T}^a,\;\;
{\cal R}_{ab}=\pm\ast{\cal R}_{ab},
\label{sd}
\end{equation}
which, however, must also satisfy a set of constraint
equations (the analogue of Einstein's equations - cf.,
(\ref{em1}) or (\ref{ym+}) in the next section).

The obvious question now is whether the above can be
{\em really} considered as a GGG, after the vierbein
$\theta^a$ has been soldered on the target space. Relatedly, 
the following arguments have appeared in the literature (the 
ennumeration is in correspondence to the one in section 1):
(i) When the vierbein $\theta^a$ is soldered onto $M^4$, our
freedom to translate is lost. (ii) Torsion is automatically absent
from the Y-M action (the ${\cal T}{\cal T}$ and ${\cal T}{\cal R}$  
contributions are easily seen to drop out, as shown in (\ref{I})). 
(iii) The absense of a linear-in-${\cal R}$ term in (\ref{I})
combined with the observation that the ${\cal R}{\cal R}$ term 
is a topological invariant (hence with no contribution to the 
classical theory) rules out reducibility to GR. 
(iv) In view of these (i-iii) impasses, the obvious resolution
is to suppress translations, set ${\cal T}=0$ (or get the same result 
from a variation of the Hilbert-Einstein action), and utilize the 
action (\ref{I}) only in the context of quantum gravity. 

We will attempt a closer examination to see that these arguments 
are misleading or just false, and then proceed to complete our 
basic results in the next section.

\section{Young-Mills action, field equations and constraints}

Still keeping the correspondence with the ennumeration (i-iv) 
in sections 1,2, we observe the following.

(i)  After soldering we indeed do not have the freedom 
to translate isometrically (or even rotate likewise), 
but all that is in $M^4$. The point is not to confuse isometries 
(and the other transformations mentioned) in $M^4$
and in $M^4_{\circ}$. To better see this, one must realize 
that although the metric on $M^4$ is identical to the metric
(\ref{m}) of $M^4_{\circ}$, the two space-times are only 
{\em locally} identical (except for any isolated
singular points in $M^4$), with the differences prescribed
by the local gauge potential ${\cal A}$. Thus, any given 
${\cal A}$ fixes the geometry of the target space 
to a particular $M^4\left({\cal A}\right)$ configuration,
generally with no symmetry at all, so that the soldered vierbein 
$\theta^a$ will necessarily be non-holonomic and the connection 
$\omega^{ab}$ non-integrable. 
$M^4$ is not even gauge invariant, because a
gauge transformed ${\cal A}'$ will generally define a
different geometry $M^4\left({\cal A}'\right)$. What {\em is}
gauge invariant, is the Y-M action (\ref{I}).

(ii) GR is not a Y-M gauge theory,
so it makes perfect sense to set there ${\cal T}=0$.
We may certainly also have configurations with ${\cal T}=0$ in 
the Y-M gauge theory of ${\cal P}$, but having {\em a priori} 
${\cal T}=0$ (equivalently, suppressing translations), simply 
means that we are dealing with the gauge theory of $SO(1,3)$, 
rather than $ISO(1,3)$. In a gauge-theoretic context, even if 
one started with $SO(1,3)$ as the gauge group, one would have 
to enlarge it to $ISO(1,3)$ (and it is in this context that one
can better appreciate Cartan's major contribution and 
foresight on the subject). The reason is that the 
representation space of $SO(1,3)$, namely $M^4_{\circ}$,
is independently also involved as locally identical to $M^4$
not in any gauge-theoretic context, but from the fundamental 
definition of $M^4$ as a pseudo-Riemannian manifold.
The nessecity to enlarge $SO(1,3)$ to $ISO(1,3)={\cal P}$ as 
a gauge group now follows from the identification of ${\cal P}$ 
and the trivial group as the isometry and holonomy groups
{\em in that order} for $M^4_{\circ}$ and {\em in reverse 
order} at the other extreme of a general $M^4$. As a result, the 
previous local identification between $M^4_{\circ}$ and $M^4$  
can be simultaneously also established in the gauge-theoretic 
context as well. This deep and elegant result would be lost 
(and GR could not be reformed to a GGG) if the torsion were 
identically zero. 

(iii) When the torsion is not identically zero,
the action (\ref{I}) {\em is not} a topological
invariant, but it is related to the
Euler characteristic of $M^4$, as we will see shortly.
As a result, variation
of (\ref{I}) with respect to $g_{ab}$ makes perfect sense.
In fact, it supplies  a well-defined energy-momentum 
tensor for the gravitational field, as well as constraints
which may be viewed as the analogue of Einstein's equations.

(iv)  With the earlier impasses out of the way, we may now 
proceed to complete our basic results.

Let ${\stackrel{\circ}{\omega}}_{ab}$, $\stackrel{\circ}{D}$,
$\stackrel{\circ}{\cal R}_{ab}$, ${\cal K}_{ab}$, denote, in 
that order, the Cristoffel part of the connection 
$\omega_{ab}={\stackrel{\circ}{\omega}}_{ab}-{\cal K}_{ab}$,
the Cristoffel covariant derivative, the associated curvature 
2-form, and the contorsion 1-form. 
Then, we may re-write (\ref{c0}) as
\begin{equation}
{\cal T}^a=D\theta^a=d\theta^a+
\left({\stackrel{\circ}{\omega}}{^a_{\cdot b}}
-{\cal K}^a_{\cdot b}\right)\wedge \theta^b=
-{\cal K}^a_{\cdot b}\wedge \theta^b,\;\;\;\;
{\cal R}_{ab}=\stackrel{\circ}{R\,}_{ab}
-{\cal H}_{ab}, 
\label{c1}
\end{equation}
where we have defined
\begin{equation}
{\cal H}^a_{\cdot b}:=\stackrel{\circ}{D}{\cal K}^a_{\cdot b}
 -{\cal K}^a_{\cdot c}\wedge{\cal K}^c_{\cdot b}.
\label{c2}
\end{equation}
The action (\ref{I}) may now equivalently be written as
\begin{equation}
{\cal I}_{YM}=\frac{1}{2}\int_{M^4}
{\cal R}^{ab}\wedge\ast{\cal R}_{ab}=
\int_{M^4}\frac{1}{2}{\stackrel{\circ}{\cal R\,}}{^{ab}}
\wedge\ast\stackrel{\circ}{\cal R}_{ab}
-{\stackrel{\circ}{\cal R\,}}{^{ab}}\wedge\ast{\cal H}_{ab}
+\frac{1}{2}
{\cal H}^{ab}\wedge\ast{\cal H}_{ab}.
\label{I1}
\end{equation}
The first term on the rhs gives the Euler 
characteristic of $M^4$, and we also observe the emergence of
the linear in ${\cal R}$ and quadratic in ${\cal T}$ terms we 
referred to earlier. 

The  energy-momentum tensor of the ${\cal F}$ field is 
determined by variation of (\ref{I1}) with respect to the 
metric $g_{ab}$ as equal to 
\begin{eqnarray}
E_{ab}:&=&\frac{1}{2}\left(R_{pqra}R^{pqr}_{\,\cdots\, b}
-\frac{1}{4}R_{pqrs}R^{pqrs}g_{ab}\right)
\nonumber \\&=&
\frac{1}{2}\left(H_{pqra}H^{pqr}_{\,\cdots \,b}
-\frac{1}{4}H_{pqrs}H^{pqrs}g_{ab}\right)
-\left(\stackrel{\circ}{R}_{pqr(a}H^{pqr}_{\,\cdots\, b)}
-\frac{1}{4}\stackrel{\circ}{R}_{pqrs}H^{pqrs}g_{ab}\right),
\label{em}
\end{eqnarray}
where $R_{\cdot\cdots}$,
$\stackrel{\circ}{R}_{\cdot\cdots}$,$H_{\cdot\cdots}$
are the components of ${\cal R}_{\cdot\cdot}$,
$\stackrel{\circ}{\cal R}_{\cdot\cdot}$,
${\cal H}_{\cdot\cdot}$ in the $\theta^a$ frame. We observe
that there are two main contributions on rhs of (\ref{em}),
with the first one clearly 
identifiable as the  energy-momentum 
tensor of the ${\cal H}$ field
\begin{equation}
E^{({\cal H})}_{ab}:=\frac{1}{2}\left(H_{pqra}H^{pqr}_{\,\cdots \,b}
-\frac{1}{4}H_{pqrs}H^{pqrs}g_{ab}\right).
\label{emh}
\end{equation}
Even if the this ${\cal H}$ 
field tuned out to be partly or entirely of non-gravitational
nature, we could subtract from (\ref{em}) that non-gravitational
contribution, to again end up with a well-defined
energy-momentum tensor for the gravitational field (cf. also
discussion in the last section).
In any case, variation of the overall action with respect to the
metric should vanish \cite{2}. Thus, if an external source 
or field were added to the action (\ref{I}), variation with 
respect to $g_{ab}$ would give
\begin{equation}
E_{ab}+E^{({\mbox sources})}_{ab}=0.
\label{em0}
\end{equation} 
This  set of ten constraints establishes the energy-momentum
balance of the theory and may be viewed as the
analogue of Einstein's equations in conventional GR, 
equivalently written as
\begin{equation}
\stackrel{\circ}{R}_{pqr(a}H^{pqr}_{\,\cdots\, b)}
-\frac{1}{4}\stackrel{\circ}{R}_{pqrs}H^{pqrs}g_{ab}=
E^{({\cal H})}_{ab}
+E^{({\mbox sources})}_{ab}.
\label{em1}
\end{equation}
We observe that these equations involve the entire Riemann
tensor (namely not just its contraction to the Ricci tensor
as in conventional GR), hence they include an explicit and
non-trivial contribution from the Weyl tensor.   
To summarize our results, the vacuum equations for our GGG are
\begin{equation}
D\ast{\cal T}^a =\ast{\cal R}^a_{\cdot b}\wedge \theta^b,
\;\;\;\;
D\ast{\cal R}^a_b=0,\;\;\;\;E_{ab}=0.
\label{ym+}
\end{equation}
The last equation does not imply zero field strength,
because ${\cal P}$ is non-compact. This will be further
discussed and seen explicity in the examples of 
the next section, to which we now turn. 

\section{Solutions and simple vacuum configurations}

There is an overlap between GR vacua and solutions 
to (\ref{ym+}), which obviouly belong to the ${\cal T}=0$
sector of the latter. One can verify
that all these solutions, common to the two theories,
fall in either of the following two classes.

Class A: All GR vacua with self-dual or anti-self-dual
Riemann tensor.

Class B: The rest of GR vacua, whose curvature satisfies
$D\ast{\cal R}^a_b=0$.

\noindent
One can prove that the Schwarzschild
black hole of GR actually belongs to
class B. This is an important result, because it
establishes for the present theory identical
Newtonian and post-Newtonian limits with GR, as
far as they are drawn from the Schwarzschild
geometry.

We will now present and examine two other solutions
which do not belong to either of the above classes.
We search for cosmological solutions to (\ref{ym+}),
with spatial homogeneity and isotropy (`Friedmann models').
This means that we may profitably choose the frame
$\theta^a$ with $\theta^0=dt$, $\theta^i=a(t)\sigma^i$
($i,j,\dots=1,2,3)$), and
with the radius (expansion scale) $a$ depending 
only on the cosmic time $t$
as indicated. The geometry of the spatial sections is
fixed by the relation 
$d\sigma^i=\frac{k}{2}\epsilon^i_{jk}\sigma^j\wedge\sigma^j$
as open (giving flat $R^3$ sections) or closed 
(giving round $S^3$ sections), depending on the value of 
the parameter $k=0,1$.

In view of the homogeneity and isotropy, the torsion cannot 
define any direction in space, so it may be choosen as
\begin{equation}
{\cal T}^a=\theta^a \wedge d\phi,\;\;\;\;\;\;\leftrightarrow
\;\;\;\;\;\;{\cal K}^{ab}=\phi^a\theta^b-\phi^b\theta^a,
\label{t}
\end{equation}
where $\phi$ is a scalar field depending only on $t$.
The system of eqs (\ref{ym+}) now admits solutions
which we have determined as
\begin{equation}
a=\frac{1}{2}\,t\,e^\phi,\;\;\;\;e^\phi
=\sqrt{\left(\frac{t}{t_0}\right)^2-k^2},
\label{a}
\end{equation}
where $t_0$ is a constant. For both $k=0,1$ values, these solutions 
represent ever-expanding universes with an initial singularity. The
expantion scale increases exactly or asymptotically as $t^2$ and 
the torsion goes to zero asymptotically as $1/t$ in the $\theta^a$ 
frame, as seen from (\ref{t}). These solutions have no GR 
counterpart: if the torsion is identically zero, then, for $k=0$ 
the expansion is lost and we only have Minkowski space-time as a 
solution to (\ref{ym+}), while for $k=1$ there is no solution at 
all.

\section{Discussion and conclusions}

We have presented here what is essentially a new theory of
gravity, based on the Y-M gauging of the Poincar\'e 
group  ${\cal P}$. Although clearly distinct from GR, it
may be viewed as a reform thereof, because it shares
with the latter its zero-torsion sector of
solutions. These we have exhaustively categorized in the 
A and B classes defined in section 4. As seen, the 
Schwarzschild black hole is a solution in class B. 
In the same section, we gave two other solutions, which
are ever-expanding Friedmann models with an initial
singularity, asymptotically zero torsion, and no GR
counterpart or limit. As mentioned, the present theory shares 
the same observational backing, and simultaneously appears 
to be free of certain pathologies of GR, as itemized below.

(i) The number of
$4\times 4+6\times 4+10=50$ {\em a priori} independent
field equations (\ref{ym+}) is
precisely equal to the number of degrees of freedom for the
likewise independent variables 
$\theta^a$,$\omega_{ab}$,$g_{ab}$. 
For the same variables, GR provides an
{\em under-determined} system of $10$ field equations plus
$24$ coming from the zero-torsion constraint \cite{4}.
We have also seen that the constraint equations (\ref{em1})
essentially replace the conventional set of Einstein's euations,
so that (unlike the case in GR), there is an explicit and
non-trivial contribution from the Weyl tensor.
  
(ii) The presence of torsion allows the elegant inter-relation 
between the gauge, isometry and general holonomy groups, as
outlined under (ii) in section 3.

(iii) There is a well-defined energy-momentum
tensor for the gravitational field. In particular, as seen 
from (\ref{em}), $E_{ab}$ is equal to zero for all solutions 
in the A or B class, in spite of the fact that the field
strength in the corresponding vaccua does not vanish (due to 
the non-compactness of ${\cal P}$, as mentioned
earlier). The same configurations viewed as GR vacua are known 
to have generally ill-defined or non-existent energy-momentum 
tensor \cite{4}.

(iv) The unification and quantization aspects of a
Y-M formulation are generally applicable in the present 
case, as long as they do not require compactness of the gauge 
group (the last qualification {\em could} of course 
re-induce some of the serious impasses associated with gravity,
but this is {\em not a priori} obvious \cite{gw}).

Expanding very briefly on the above (and perhapse also 
on the side of speculation), 
we note, on the one hand, the apparent association of
$\omega_{ab}$ (or $\stackrel{\circ}{\omega}_{ab}$) with what 
should be cosidered as the gravitational bosons. At the same 
time, with the linear-in-${\cal R}$ contribution 
in the action (\ref{I1}) 
viewed as a generalization of the Hilbert-Einstein Lagrangian,
it is clear from (\ref{em1}) that the generally non-constant 
gravitational coupling is inherently 
specified, e.g., as function of $\phi$ 
in the simpler models. In a closely related `Machian' 
behavior, such couplings have been quite generally shown
to be positive and asymptotically constant \cite{b1}. 
On the other hand, the physical identification
of the vierbein $\theta^a$ as a gauge field remains unclear,
obviously dependent on the physical identification of
the tosrsion and consequently of the ${\cal H}$ field in 
the action (\ref{I1}). If not of gravitational
nature, the ${\cal H}$ field could be the carrier of some 
other fundamental interaction (e.g., as exemplified by 
the $\phi$ field in the simple
models of section 4). In that context, the electroweak 
interaction appears to be a plaussible candidate
(e.g., as attempted in recent work \cite{b2}).
In any case, torsion is here a dynamical (`propagating') 
field, so its direct association with
spin \cite{2} is ruled out. 

\vspace{.5cm}
I am grateful to A.A. Kehagias for discussions.

\end{document}